\documentclass{revtex4}
\usepackage{amsthm,amssymb,amsmath}				
\usepackage{graphicx,comment}
\usepackage{float}   
 \usepackage{xcolor}

\begin{document}

\title{$n$-dimensional PDM-damped harmonic oscillators: Linearizability, and exact
solvability}
\author{Omar Mustafa}
\email{omar.mustafa@emu.edu.tr}
\affiliation{Department of Physics, Eastern Mediterranean University, G. Magusa, north
Cyprus, Mersin 10 - Turkey.}

\begin{abstract}
\textbf{Abstract:}\ We consider position-dependent mass (PDM) Lagrangians/Hamiltonians in their standard textbook form, where the long-standing \emph{gain-loss balance} between the kinetic and potential energies is kept intact to allow conservation of total energy (i.e., $L=T-V$, $H=T+V$, and $dH/dt=dE/dt=0$). Under such standard settings, we discuss and report on $n$-dimensional PDM damped harmonic oscillators (DHO). We use some $n$-dimensional point canonical transformation to facilitate the linearizability of their $n$-PDM dynamical equations into some $n$-linear DHOs' dynamical equations for constant mass setting. Consequently, the well know exact solutions for the linear DHOs are mapped, with ease, onto the exact solutions for PDM DHOs. A set of one-dimensional and a set of $n$-dimensional PDM-DHO illustrative examples are reported along with their phase-space trajectories.


\end{abstract}

\maketitle

\section{Introduction}

A large number of research on classical and quantum particles endowed with \emph{"effective"} position-dependent mass (PDM) has been carried out [1-48] ever since the introduction of the Mathews-Lakshmanan (M-L) oscillator \cite{M-L 1974}. Such PDM concept manifestly introduces a mathematically challenging problem in both classical and quantum mechanics. On the quantum mechanical side, the PDM von Roos Hamiltonian \cite{von Roos} is indulged with a parametric ordering ambiguity problem, infected by the non-unique
representation of the said Hamiltonian (c.f., e.g. \cite{Mustafa Habib 2007,Bagchi 2004,Cruz 2007,Mustafa 2020} and references cited therein). Only very recently that a proper definition for the position-dependent mass momentum operator is introduced by Mustafa and Algadhi \cite{Mustafa Algadhi 2019}, resolving hereby the ordering ambiguity problem. Moreover, using the von Roos Hamiltonian \cite{von Roos}, Mustafa \cite{Mustafa 2020} has shown that the traditional textbook commutation relation between the PDM creation $%
\hat{A}^{+}$ and annihilation $\hat{A}$ operators, $\left[ \hat{A},\hat{A}^{+}\right] =1$, offers a unique parametric ordering (known in the literature as MM-ordering \cite{Mustafa Habib 2007}) for the PDM von Roos Hamiltonian. On the classical mechanical side, however, the mean stream of research focuses on a set of interesting non-standard/non-traditional PDM Lagrangian/Hamiltonian settings \cite{M-L 1974,Venk laksh 1997,Tiwari 2013,Lak-Chand 2013,Pradeep 2009,Chand-Lak 2007,Carl Bender 2016,Musielak 2008,Bhuvaneswari 2012,Carinena Ranada 2005,Carinena Ranada Sant
2005,Carinena Ranada Sant 2004,Carinena Sigma 2007,Rodrigues 2008,Mustafa Phys.Scr. 2020,Mustafa 2015,Ranada 2016,Carinena Herranz 2017,Bagchi Ghosh 2013,Mustafa 2013} , where the intimate long-standing\emph{\ gain-loss balance} between the kinetic and potential energies is ignored in the process, and consequently the total energy of the dynamical system is no longer a conserved quantity \cite{Carl Bender 2016,Mustafa 2020}.

The position-dependent mass terminology, nevertheless, should be understood as a metaphorical consequential manifestation of some position-dependent deformation in the coordinate system or, equivalently, a position-dependent deformation in the velocity of the particle at hand. Yet, it has been reported that a point mass moving within the curved coordinates/space transforms, effectively, into a PDM in Euclidean coordinates/space (c.f., e.g., \cite{Carinena Ranada Sant 2004,Mustafa 2015,Mustafa 2019,Khlevniuk 2018,Mustafa Algadhi 2019} and references cited therein). This understanding leads, in fact, to an interesting set of standard/traditional PDM
Lagrangians/Hamiltonians with their exact solutions are inferred from the well known exactly solvable dynamical systems (e.g., Mustafa \cite{Mustafa arXiv}). That is, a standard Lagrangian/Hamiltonian (i.e., $L=T-V$, $H=T+V$) in the generalized coordinates%
\begin{equation}
L\left( q,\dot{q},t\right) =\frac{1}{2}m_{\circ }\dot{q}^{2}-V\left(
q\right) \Longleftrightarrow H\left( q,p_{q},t\right) =\frac{p_{q}^{2}}{%
2m_{\circ }}+V\left( q\right) ;\,p_{q}=m_{\circ }\dot{q}
\end{equation}%
would transform into%
\begin{equation}
L\left( x,\dot{x},t\right) =\frac{1}{2}m_{\circ }m\left( x\right) \dot{x}%
^{2}-V\left( q(x)\right) \Longleftrightarrow H\left( x,p_{x},t\right) =\frac{%
p_{x}^{2}}{2m_{\circ }m\left( x\right) }+V\left( q(x)\right)
;\,p_{x}=m\left( x\right) \dot{x},
\end{equation}%
under the point canonical transformation%
\begin{equation}
q\longrightarrow q\left( x\right) =\int \sqrt{m\left( x\right) }dx=\sqrt{%
Q\left( x\right) }x\Longleftrightarrow \sqrt{m\left( x\right) }=\sqrt{%
Q\left( x\right) }\left( 1+\frac{Q^{\prime }\left( x\right) }{2Q\left(
x\right) }x\right) .  \label{1D PT}
\end{equation}%
This would in turn yield the Euler-Lagrange dynamical invariance%
\begin{equation}
\frac{d}{dt}\left( \frac{\partial L}{\partial \dot{q}}\right) -\frac{%
\partial L}{\partial q}=0=\frac{d}{dt}\left( \frac{\partial L}{\partial \dot{%
x}}\right) -\frac{\partial L}{\partial x}
\end{equation}%
between the two dynamical systems. For example, the harmonic oscillator Lagrangian%
\begin{equation}
L\left( q,\dot{q},t\right) =\frac{1}{2}m_{\circ }\dot{q}^{2}-\frac{1}{2}%
m_{\circ }\omega ^{2}q^{2}\Longleftrightarrow L\left( x,\dot{x},t\right) =%
\frac{1}{2}m_{\circ }m\left( x\right) \dot{x}^{2}-\frac{1}{2}m_{\circ
}\omega ^{2}Q\left( x\right) x^{2}
\end{equation}%
would yield the Euler-Lagrange dynamical invariance%
\begin{equation}
\ddot{q}+\omega ^{2}q=0=\ddot{x}+\frac{m^{\prime }\left( x\right) }{2m\left(
x\right) }\dot{x}^{2}+\sqrt{\frac{Q\left( x\right) }{m\left( x\right) }}%
\omega ^{2}x
\end{equation}%
where%
\begin{equation}
q=A\cos \left( \omega t+\varphi \right) =\sqrt{Q\left( x\right) }x.
\end{equation}%
Moreover, the long-standing \emph{gain-loss balance} between the kinetic and potential energies is kept intact (through the correlation in (3) between the two dimensionless scalar multipliers $%
\sqrt{m\left( x\right) }$ and $\sqrt{Q\left( x\right) }$). This would, in turn, facilitate exact solvability of the dynamical system at hand. For more
details on this issue the reader may refer to Mustafa \cite{Mustafa arXiv} for illustrative examples on isochronuous PDM oscillators in one- and $n$-dimensions. Moreover, equation (6) offers a straightforward linearization of the quadratic Li\'{e}nard-type differential equation $\ddot{x}+f\left(
x\right) \dot{x}^{2}+g\left( x\right) =0$, with  $f\left( x\right)
=m^{\prime }\left( x\right) /2m\left( x\right) $ and $g\left( x\right) =%
\sqrt{Q\left( x\right) /m\left( x\right) }\omega ^{2}x$, into a simple linear oscillator $\ddot{q}+\omega ^{2}q=0$ and inherits its exact solution. 

It would be interesting, therefore, to study PDM classical particles moving under the influence of a conservative oscillator potential force field $%
V\left( q\right) =m_{\circ }\omega ^{2}q^{2}/2$ and a non-conservative Rayleigh dissipative force field $\mathcal{R}\left( \dot{q}\right) =b\,\dot{q%
}^{2}$ \cite{da Costa1 2020} (an elegant and convenient tool to accommodate dissipative forces) and described by the damped harmonic oscillator (DHO) dynamical equation%
\begin{equation}
\frac{d}{dt}\left( \frac{\partial L}{\partial \dot{q}}\right) -\frac{%
\partial L}{\partial q}+\frac{\partial \mathcal{R}}{\partial \dot{q}}%
=0\Longleftrightarrow \ddot{q}+2\eta \omega \,\dot{q}+\omega ^{2}q=0;\,\eta =%
\frac{b}{2m_{\circ }\omega },  \label{DHO}
\end{equation}%
where $\eta >0$ is the damping ratio and $\eta =0$ corresponds to the linear harmonic oscillator. Apparently, this equation of motion is a linear Li\'{e}nard-type differential equation (i.e., $\ddot{q}+h\left( q\right) \dot{q}%
+g\left( q\right) =0$, $h\left( q\right) =2\eta \omega \,$\ and $g\left(
q\right) =\omega ^{2}q$) that admits an exact solution of the form%
\begin{equation}
q\left( t\right) =e^{-\omega \eta t}\left[ A\cosh \left( \beta t\right)
+B\sinh \left( \beta t\right) \right] \,;\,\beta =\omega \sqrt{\eta ^{2}-1},
\label{DHO solution}
\end{equation}%
and subjected to the initial values of the problem at hand. Moreover, it is worthwhile to mention that the values of the damping ratio $\eta $ determine the nature of damping. Namely, $\eta >1$ addresses "over-damping", $\eta =1$ addresses "critical damping", and $\eta <1$ addresses "under-damping". Nevertheless, it is obvious that the DHO equation of (\ref{DHO}) cannot be obtained from a standard Hamiltonian with a conserved total energy $E=\frac{1%
}{2}m_{\circ }\dot{q}^{2}+\frac{1}{2}m_{\circ }\omega ^{2}q^{2}$.  The DHO equation (\ref{DHO}) readily indicates that%
\begin{equation}
\frac{dE}{dt}=\frac{d}{dt}\left( \frac{1}{2}m_{\circ }\dot{q}^{2}+\frac{1}{2}%
m_{\circ }\omega ^{2}q^{2}\right) =-2\eta \omega \,\dot{q}^{2}\neq
0;\,\forall \eta >0,  \label{energy conservation}
\end{equation}%
where the total energy decreases as the dynamical system evolves in time \cite{Pradeep 2009,Chand-Lak 2007,Carl Bender 2016}. Yet, unlike the conservative dynamical systems where the phase space remains constant, the phase space of a dissipative non-conservative dynamical systems tends to decrease/shrink with time. Moreover, the DHO in (\ref{DHO}) admits invariance under our point canonical transformation recipe (\ref{1D PT}) to imply PDM damped harmonic oscillators%
\begin{equation}
\frac{d}{dt}\left( \frac{\partial L}{\partial \dot{x}}\right) -\frac{%
\partial L}{\partial x}+\frac{\partial \mathcal{R}}{\partial \dot{x}}%
=0\Longleftrightarrow \ddot{x}+\frac{m^{\prime }\left( x\right) }{2m\left(
x\right) }\dot{x}^{2}+2\eta \omega _{\circ }\dot{x}+\sqrt{\frac{Q\left(
x\right) }{m\left( x\right) }}\omega _{\circ }^{2}x=0.  \label{PDM-DHO}
\end{equation}%
This should, in turn, not only document a linearization of a mixed Li\'{e}nard-type differential equation (i.e., $\ddot{x}+f\left( x\right) \dot{x}%
^{2}+h\left( x\right) \dot{x}+g\left( x\right) =0;$ $f\left( x\right)
=m^{\prime }\left( x\right) /2m\left( x\right) ,h\left( x\right) =2\eta
\omega _{\circ }$ and $g\left( x\right) =\sqrt{Q\left( x\right) /m\left(
x\right) }\omega ^{2}x$) into a linear Li\'{e}nard-type differential equation (i.e., $\ddot{q}+\tilde{h}\left( q\right) \dot{q}+\tilde{g}\left( q\right) =0$, $\tilde{h}\left( q\right) =2\eta \omega$ and $\tilde{g}\left( q\right) =\omega ^{2}q$), but also give the corresponding PDM-deformed DHO's dynamical system that inherits the exact solvability of the DHO in (\ref{DHO solution}) (to be exemplified in the current methodical proposal below). To the best of our knowledge, such methodical proposal/approach has never been reported elsewhere.


In the current study, we consider (in section II) the $n$-dimensional PDM damped harmonic oscillators within the standard textbook format, where the long-standing \emph{gain-loss balance} between the kinetic and potential energies is kept intact to allow conservation of total energy (i.e., $L=T-V$, $H=T+V$, and $dH/dt=dE/dt=0$). Therein, we use some $n$-dimensional point canonical transformation to facilitate the linearizability of their $n$-PDM dynamical equations into some $n$-linear DHOs' dynamical equations for constant mass setting. Hence, the well know exact solutions for the linear DHOs are mapped, with ease, onto the exact solutions for PDM DHOs. A set of one-dimensional and a set of $n$-dimensional PDM-DHO illustrative examples are reported, along with their phase-space trajectories, in section III.  Our concluding remarks are given in section IV.

\section{PDM damped harmonic oscillators in $n$-dimensions}

Consider a classical particle of mass $m_{\circ }$ moving, within the generalized $n$-dimensional\ coordinates, under the influence of the $n$-dimensional harmonic oscillator potential $V\left( \mathbf{q}\right) $ (i.e., conservative force field) and the $n$-dimensional Rayleigh dissipative force field $\mathcal{R}\left( \dot{q}\right) $ (i.e., non-conservative force field) given, respectively, by%
\begin{equation}
V\left( \mathbf{q}\right) =\frac{1}{2}m_{\circ }\omega _{\circ
}^{2}\sum\limits_{j=1}^{n}q_{_{j}}^{2}\,,  \label{V(q)}
\end{equation}%
and%
\begin{equation}
\,\mathcal{R}\left( \mathbf{\dot{q}}\right) =b\sum\limits_{j=1}^{n}\dot{q}%
_{_{j}}^{2},  \label{G(q)}
\end{equation}%
where $\omega _{\circ }=\sqrt{k/m_{\circ }}$ is the undamped angular frequency and $b$ is the damping coefficient. Then, the corresponding Lagrangian reads%
\begin{equation}
L\left( \mathbf{q,\dot{q},t}\right) =\frac{1}{2}m_{\circ
}\sum\limits_{j=1}^{n}\dot{q}_{_{j}}^{2}-\frac{1}{2}m_{\circ }\omega _{\circ
}^{2}\sum\limits_{j=1}^{n}q_{_{j}}^{2},  \label{q-Lagrangian}
\end{equation}%
and the $n$-Euler-Lagrange equations of motion become%
\begin{equation}
\frac{d}{dt}\left( \frac{\partial L}{\partial \dot{q}_{i}}\right) -\frac{%
\partial L}{\partial q_{i}}+\frac{\partial \mathcal{R}}{\partial \dot{q}_{i}}%
=0;\text{ }\,\,\,i=1,2,\cdots ,n\in 
\mathbb{N}
.  \label{EL-G}
\end{equation}%
Consequently, in a straightforward manner, one obtains the $n$-equations of motion%
\begin{equation}
\ddot{q}_{i}+2\eta \omega _{\circ }\,\dot{q}_{i}+\omega _{\circ
}^{2}q_{i}=0;\,\eta =\frac{b}{2{m_{\circ }\omega _{\circ }}},i=1,2,\cdots ,n\in 
\mathbb{N}
,  \label{q-equation of motion}
\end{equation}%
which are known as the damped harmonic oscillators equations of motion, where $\eta$ identifies the damping ratio and the second term represents the effect of damping. In the literature, moreover, this equation is also known to belong to the class of the linear-type Li\'{e}nard oscillators. Equation (\ref{q-equation of motion}) on the other hand, admits exact solutions in the forms of%
\begin{equation}
q_{i}\left( t\right) =e^{-\eta \omega _{\circ }t}\left[ A_{i}\cosh \left( \beta
t\right) +B_{i}\sinh \left( \beta t\right) \right] ;\,\beta =\omega _{\circ }%
\sqrt{\eta ^{2}-1}.  \label{q-solution}
\end{equation}%
This solution is to satisfy and suit the initial boundary conditions, of course. Throughout the current proposal, we require that $q_{i}\left(
0\right) \neq 0$ and $\dot{q}_{i}\left( 0\right) =0$ so that the solutions in (\ref{q-solution}) reduce to%
\begin{equation}
q_{i}\left( t\right) =A_{i}e^{-\eta \omega _{\circ }t}\cosh \left( \beta
t\right) ;\,\beta =\omega _{\circ }\sqrt{\eta ^{2}-1}.  \label{q-adopted}
\end{equation}%
Obviously, the values of the damping ratio $\eta $ in (\ref{q-adopted}) determine the nature/behavior of the solution and suggest the following classifications of the system. That is, for $\eta >1$ the system is classified as an \emph{"Over-damped system" }that exponentially decays to its steady/equilibrium state without oscillating (the larger the values of $\eta $ the slower the decay is). For $\eta =1$ the \ system is a \emph{"Critically-damped system"} \ and decays to its steady/equilibrium state as fast as possible. Finally, for $\eta <1$ the system is an \emph{"Under-damped system"}  and oscillates with the amplitude gradually decreases to zero.

We may now follow the canonical point transformation introduced by Mustafa \cite{Mustafa 2019,Mustafa 2020,Mustafa arXiv} and define%
\begin{equation}
dq_{_{i}}=\delta _{ij}\sqrt{m\left( \mathbf{r}\right) }\,dx_{_{j}}%
\Longleftrightarrow \frac{\partial q_{_{i}}}{\partial x_{_{j}}}=\delta _{ij}%
\sqrt{m\left( \mathbf{r}\right) }\Longleftrightarrow q_{_{i}}\left( \mathbf{r%
}\right) =\int \sqrt{m\left( \mathbf{r}\right) }\,dx_{i}=\sqrt{Q\left( 
\mathbf{r}\right) }x_{i}.  \label{point-transformation1}
\end{equation}%
This would, with $m\left( \mathbf{r}\right) =m\left(
r\right) $, $Q\left( \mathbf{r}\right) =Q\left( r\right) $ (to avoid complexity of calculations) and%
$$\hat{q}_{_{i}}=\frac{\sum\limits_{k=1}^{n}\left( \frac{\partial x_{_{k}}}{%
\partial q_{_{i}}}\right) \hat{x}_{k}}{\sqrt{\sum\limits_{k=1}^{n}\left( 
\frac{\partial x_{_{k}}}{\partial q_{_{i}}}\right) ^{2}}}\Longleftrightarrow 
\hat{q}_{_{i}}=\,\hat{x}_{_{i}}$$%
immediately imply, using (\ref{point-transformation1}) along with $\hat{q}_{_{i}}=\hat{x}_{_{i}}$, that%
\begin{equation}
\mathbf{q}\left( \mathbf{r%
}\right) \mathbf{=}\sqrt{Q\left( r\right) }\mathbf{r},
\label{point-transformation2}
\end{equation}%
and, by using (\ref{point-transformation1}),%
\begin{equation}
\dot{q}_{_{i}}\left( \mathbf{r}\right) =\sum\limits_{k=1}^{n}\partial
_{x_{k}}q_{_{i}}\left( \mathbf{r}\right) \,\dot{x}_{_{k}}=\dot{x}_{_{i}}%
\sqrt{m\left( r\right) }\Longleftrightarrow \mathbf{\dot{q}}\left( \mathbf{r}%
\right) =\sqrt{m\left( r\right) } \mathbf{\dot{r}}.
\label{velocity-transformation}
\end{equation}%
It is obvious that the relation between $m\left( r\right) $ and $Q\left(
r\right) $ is determined, in a straightforward manner, by (\ref{point-transformation2}) and (\ref{velocity-transformation}) as%
\begin{equation}
\sqrt{m\left( r\right) }=\sqrt{Q\left( r\right) }\left( 1+\frac{Q^{\prime
}\left( r\right) }{2Q\left( r\right) }r\right) ;\,r=\sqrt{%
\sum\limits_{j=1}^{n}x_{_{j}}^{2}}.  \label{Q-m relation}
\end{equation}%
Here, we have used the assumption that $\mathbf{r}\parallel \mathbf{\dot{r}%
\Longleftrightarrow }\left( \mathbf{\dot{r}\cdot r}\right) \mathbf{r=}\left( 
\mathbf{r\cdot r}\right) \mathbf{\dot{r}=}r^{2}\mathbf{\dot{r}}$ (i.e., no rotational effects are involved in the problem at hand). Consequently, the $n
$ Euler-Lagrange equations (\ref{EL-G}) would result%
\begin{equation}
\ddot{x}_{_{i}}+\frac{m^{\prime }\left( r\right) }{2rm\left( r\right) }%
\left( \mathbf{\dot{r}\cdot r}\right) \dot{x}_{_{i}}+2\eta \omega _{\circ }%
\dot{x}_{_{i}}+\sqrt{\frac{Q\left( r\right) }{m\left( r\right) }}\omega
_{\circ }^{2}x_{i}=0.  \label{EL-transformed}
\end{equation}%
On the other hand, let us consider a PDM-particle moving in a conservative PDM-deformed oscillator potential force field%
\begin{equation}
V\left( \mathbf{r}\right) =\frac{1}{2}m_{\circ }\omega _{\circ }^{2}Q\left(
r\right) \sum\limits_{j=1}^{n}x_{_{j}}^{2}  \label{PDM-oscillator}
\end{equation}%
and subjected to a non-conservative PDM Rayleigh dissipation function%
\begin{equation}
\mathcal{R}\left( \mathbf{r},\mathbf{\dot{r}}\right) =b m\left( r\right)
\sum\limits_{j=1}^{n}\dot{x}_{_{j}}^{2}.  \label{PDM-Rayleigh}
\end{equation}%
Then the standard PDM-Lagrangian describing this particle reads%
\begin{equation}
L\left( \mathbf{r,\dot{r},t}\right) =\frac{1}{2}m_{\circ }m\left( r\right)
\sum\limits_{j=1}^{n}\dot{x}_{_{j}}^{2}-\frac{1}{2}m_{\circ }\omega _{\circ
}^{2}Q\left( r\right) \sum\limits_{j=1}^{n}x_{_{j}}^{2},
\label{PDM-Lagrangian}
\end{equation}%
and the $n$-PDM Euler-Lagrange equations are given by%
\begin{equation}
\frac{d}{dt}\left( \frac{\partial L}{\partial \dot{x}_{i}}\right) -\frac{%
\partial L}{\partial x_{i}}+\frac{\partial \mathcal{R}}{\partial \dot{x}_{i}}%
=0  \label{EL-dissipative}
\end{equation}%
to imply the $n$-PDM equations of motion%
\begin{equation}
\ddot{x}_{i}+\frac{m^{\prime }\left( r\right) }{rm\left( r\right) }\left( 
\mathbf{\dot{r}\cdot r}\right) \dot{x}_{i}-\frac{m^{\prime }\left( r\right) 
}{2rm\left( r\right) }\left( \mathbf{\dot{r}\cdot \dot{r}}\right)
x_{i}+2\eta \omega _{\circ }\dot{x}_{i}+\sqrt{\frac{Q\left( r\right) }{%
m\left( r\right) }}\omega _{\circ }^{2}x_{i}=0.  \label{PDM-EL}
\end{equation}%
Apparently, the Euler-Lagrange equations of (\ref{EL-transformed}) (manifestly introduced by the canonical point transformation (\ref{point-transformation1})) and the PDM ones of (\ref{PDM-EL}) are still far from securing invariance. We, therefore, appeal to the recently introduced total vector \emph{Newtonian invariance amendment}, i.e., %
\begin{equation}
\sum_{i=1}^{n}\left( \frac{d}{dt}\left( \frac{\partial L}{\partial \dot{q}%
_{i}}\right) -\frac{\partial L}{\partial q_{i}}+\frac{\partial \mathcal{R}}{%
\partial \dot{q}_{i}}\right) \hat{x}_{i}=0=\sum_{i=1}^{n}\left( \frac{d}{dt}%
\left( \frac{\partial L}{\partial \dot{x}_{i}}\right) -\frac{\partial L}{%
\partial x_{i}}+\frac{\partial \mathcal{R}}{\partial \dot{x}_{i}}\right) 
\hat{x}_{i}  \label{Newtonian invariance}
\end{equation}%
by Mustafa \cite{Mustafa Phys.Scr. 2020,Mustafa arXiv}, and multiply both equations (\ref{EL-transformed}) and (\ref{PDM-EL}), by the corresponding unit vectors $\hat{x}_{_{i}}$ and sum over $i=1,2,\cdots ,n$ \ to obtain (from both equations)%
\begin{equation}
\mathbf{a}+\frac{m^{\prime }\left( r\right) }{2rm\left( r\right) }v^{2}%
\mathbf{r}+2\eta \omega _{\circ }\mathbf{\dot{r}}+\sqrt{\frac{Q\left( r\right) }{%
m\left( r\right) }}\omega _{\circ }^{2}\mathbf{r}=0,  \label{Newtonian equations}
\end{equation}%
where $$v^{2}=\sum\limits_{j=1}^{n}\dot{x}_{_{j}}^{2},  \,\mathbf{r=}\sum\limits_{j=1}^{n}x_{_{j}}\hat{x}_{_{j}}.$$At this point, however, it is obvious that equation (\ref{Newtonian invariance}) is nothings but the total vector presentation of the Euler-Lagrange vector-component dynamical equations. In fact, this procedure identifies the textbook mapping between Euler-Lagrange and Newtonian dynamics. Moreover, both equations, (\ref{EL-transformed}) and (\ref{PDM-EL}), lead to the same Newtonian dynamical equation (\ref{Newtonian equations}). Hence, we may now safely decompose equation (\ref{Newtonian equations}) into its components to read%
\begin{equation}
\ddot{x}_{i}+\frac{m^{\prime }\left( r\right) }{2rm\left( r\right) }%
x_{i}\left( \sum\limits_{j=1}^{n}\dot{x}_{_{j}}^{2}\right) +2\eta \omega
_{\circ }\dot{x}_{i}+\sqrt{\frac{Q\left( r\right) }{m\left( r\right) }}%
\omega _{\circ }^{2}x_{i}=0;\text{ }\,\,\,i=1,2,\cdots ,n\in 
\mathbb{N}
.  \label{PDM damped oscillator}
\end{equation}%
These equations represent the equations of motion for a PDM damped oscillator described by the standard $n$-dimensional PDM-Lagrangian (\ref{PDM-Lagrangian}) for a PDM-particle subjected to a PDM Rayleigh dissipative force field (\ref{PDM-Rayleigh}).

Furthermore, it is unavoidable to mention the manifestly user friendly transition procedure between the Li\'{e}nard-type differential equations. Namely, in the one-dimension, equation (\ref{PDM damped oscillator}) represent a subset of the mixed-type Li\'{e}nard equation%
\begin{equation}
\ddot{x}+f\left( x\right) \,\dot{x}^{2}+h\left( x\right) \,\dot{x}+g\left(
x\right) =0  \label{mixed-type Lienard equation}
\end{equation}
where%
\begin{equation}
f\left( x\right) =\frac{m^{\prime }\left( x\right) }{2m\left( x\right) }%
,\,h\left( x\right) =2\eta \omega _{\circ },\,g\left( x\right) =\sqrt{\frac{%
Q\left( x\right) }{m\left( x\right) }}\omega _{\circ }^{2}x. \label{PDM-type Lienard equation}
\end{equation}%
Which admits linearizability into the one dimensional linear Li\'{e}nard-type oscillator (\ref{q-equation of motion}) (i.e., the damped harmonic oscillator equation) through some reverse engineering of our canonical point transformation above. Yet, for $\eta =0$ equation (\ref{PDM damped oscillator}) reads the quadratic Li\'{e}nard-type oscillator which is obviously linearizable into a simple harmonic oscillator equation $\ddot{q}%
+\omega _{\circ }^{2}q=0;\,q=A\cos \left( \omega _{\circ }t+\varphi \right) $ .

Under such PDM settings, one would use (\ref{PDM-Lagrangian}) and obtain%
\begin{equation}
p_{i}=\frac{\partial L}{\partial \dot{x}_{i}}=m(r)\dot{x}_{i}\Longleftrightarrow \dot{p}_{i}=%
\frac{d}{dt}\left( \frac{\partial L}{\partial \dot{x}_{i}}\right) =\frac{%
\partial L}{\partial x_{i}}-\frac{\partial \mathcal{R}}{\partial \dot{x}_{i}}%
.  \label{dynamical variables}
\end{equation}%
In what follows we discuss some illustrative examples in one- and $n$-dimensions.%
\begin{figure}[h!]  
\centering
\includegraphics[width=0.3\textwidth]{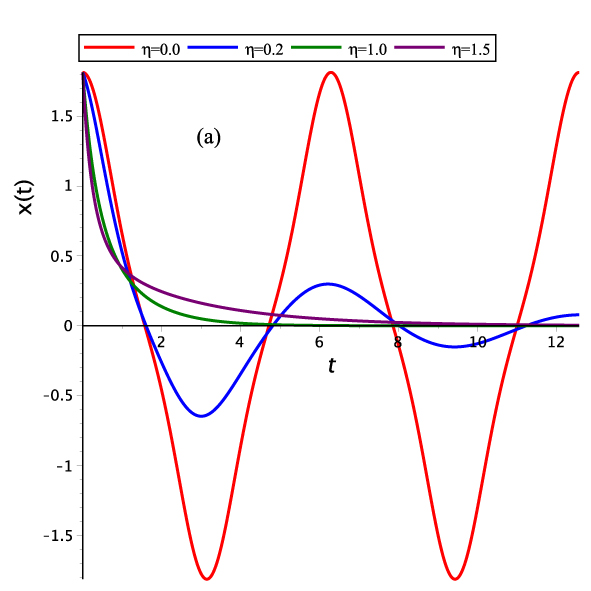} 
\includegraphics[width=0.3\textwidth]{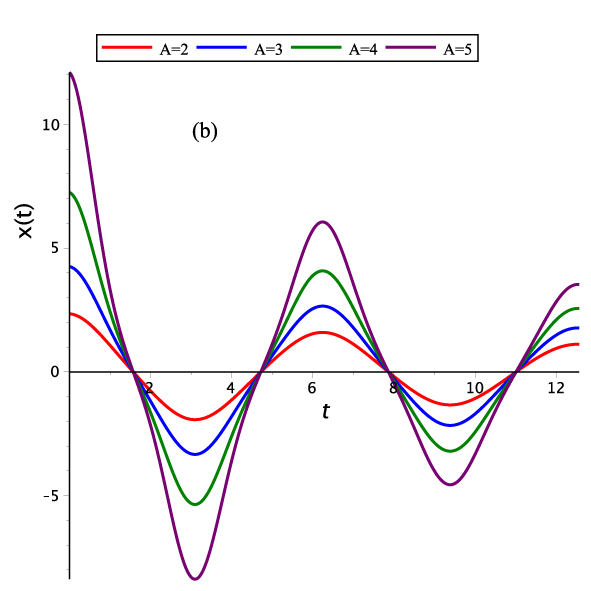}
\includegraphics[width=0.3\textwidth]{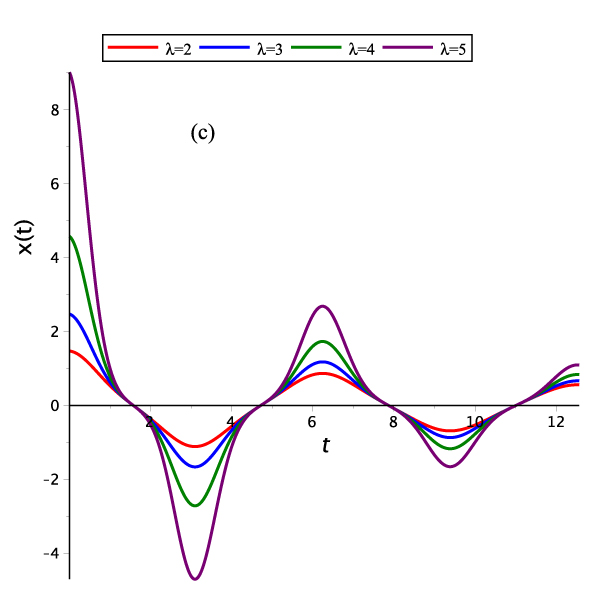}
\includegraphics[width=0.3\textwidth]{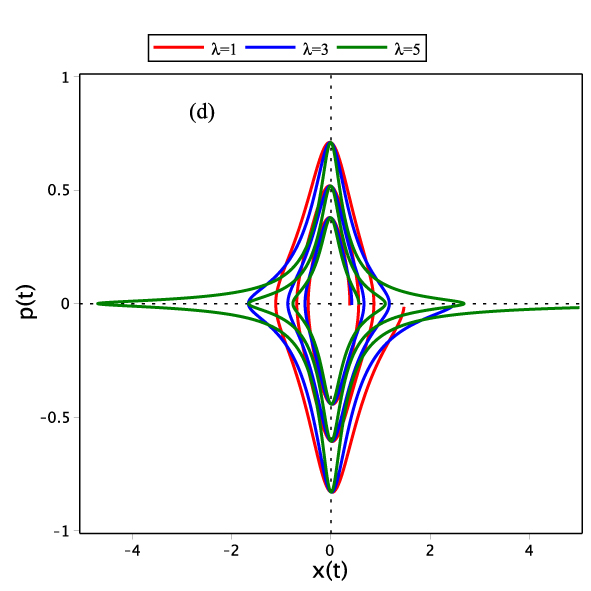}
\includegraphics[width=0.3\textwidth]{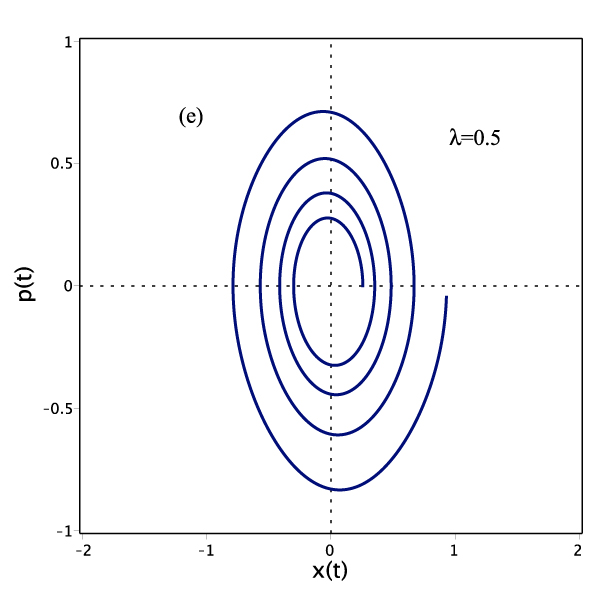}
\includegraphics[width=0.3\textwidth]{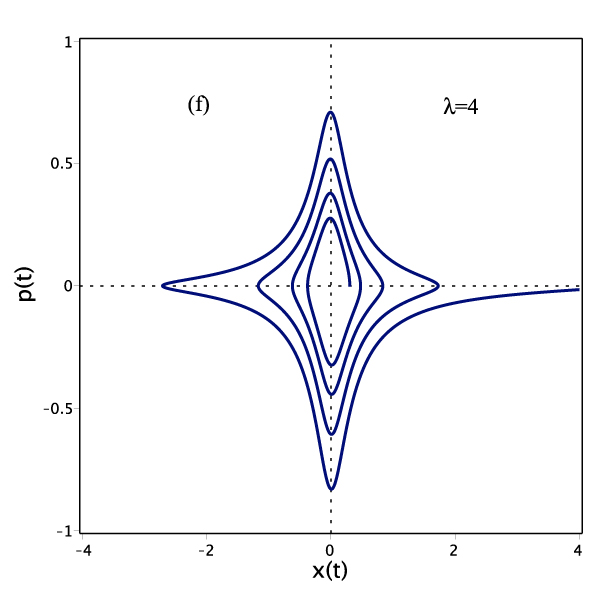}
\caption{\small 
{ For the PDM $m(x)$ in (\ref{ML-PDM}), we show the effect of (a) the damping ratio $\eta$ (at $\omega=1, \lambda=2, A=1$), (b) the amplitude $A$ (at $\omega=1, \lambda=0.5, \eta=0.05$, under-damped), and (c) the PDM parameter $\lambda$ (at $\omega=1, A=0.9, \eta=0.05$, under-damped), on the coordinate $x(t)$ of (\ref{x(t)-MLPDM}) as it evolves in time. The phase trajectories $p(t)$ vs $x(t)$ (at $\omega=1, A=0.9, \eta=0.05$, under-damped) are shown in (d) for $\lambda=1, 2, 3$, (e) $\lambda=0.5$, and (f) $\lambda=4$.
}}
\label{fig5}
\end{figure}

\section{Illustrative examples}

\subsection{One-dimensional PDM damped oscillators:}

\subsubsection{A velocity deformation $m\left( x\right) =1/\left( 1+\lambda ^{2}x^{2}\right) $}

A PDM of the form%
\begin{equation}
m\left( x\right) =\frac{1}{1+\lambda ^{2}x^{2}}  \label{ML-PDM}
\end{equation}%
would lead, through (\ref{Q-m relation}), to%
\begin{equation}
\sqrt{Q\left( x\right) }x=\frac{1}{\lambda }\ln \left( \lambda x+\sqrt{%
1+\lambda ^{2}x^{2}}\right) .  \label{Q(x)-ML}
\end{equation}%
Using (\ref{point-transformation1}) and (\ref{q-adopted}), one would write%
\begin{equation}
q\left( x\right) =Ae^{-\omega _{\circ }\eta t}\cosh \left( \beta t\right) =%
\frac{1}{\lambda }\ln \left( \lambda x+\sqrt{1+\lambda ^{2}x^{2}}\right) ,
\end{equation}%
to imply%
\begin{equation}
x\left( t\right) =\frac{1}{\lambda }\sinh \left( A\lambda e^{-\omega _{\circ
}\eta t}\cosh \left( \beta t\right) \right) ,  \label{x(t)-MLPDM}
\end{equation}%
as an exact solution for the one-dimensional form of (\ref{EL-transformed}) that reads the dynamical equation of motion%
\begin{equation}
\ddot{x}-\frac{\lambda^2 x}{1+\lambda ^{2}x^{2}}\dot{x}^{2}+2\eta \omega
_{\circ }\dot{x}+\frac{\omega _{\circ }^{2}}{\lambda }\sqrt{1+\lambda
^{2}x^{2}}\ln \left( \lambda x+\sqrt{1+\lambda ^{2}x^{2}}\right) =0.
\label{ML-m(x) equation of motion}
\end{equation}%
This equation of motion corresponds to the PDM-particle (\ref{ML-PDM}) moving in the vicinity of a conservative potential force field%
\begin{equation}
V\left( x\right) =\frac{1}{2}\omega _{\circ }^{2}\left( \frac{1}{\lambda }%
\ln \left( \lambda x+\sqrt{1+\lambda ^{2}x^{2}}\right) \right) ^{2},
\label{ML-potential}
\end{equation}%
and feels a non-conservative Rayleigh dissipative force field%
\begin{equation}
\mathcal{R}\left( x,\dot{x}\right) =\frac{b}{1+\lambda ^{2}x^{2}}\dot{x}^{2}.
\label{ML-dissipative potential}
\end{equation}%
Moreover, the corresponding PDM-standard Lagrangian and PDM-Hamiltonian are  given by%
\begin{equation}
L=\frac{1}{2}\frac{\dot{x}^{2}}{1+\lambda ^{2}x^{2}}-\frac{1}{2}\omega
_{\circ }^{2}\left( \frac{1}{\lambda }\ln \left( \lambda x+\sqrt{1+\lambda
^{2}x^{2}}\right) \right) ^{2},  \label{ML-Lagrangian}
\end{equation}%
and%
\begin{equation}
H=(1+\lambda ^{2}x^{2})\frac{p_{_{x}}^2}{2} +\frac{1}{2}\omega
_{\circ }^{2}\left( \frac{1}{\lambda }\ln \left( \lambda x+\sqrt{1+\lambda
^{2}x^{2}}\right) \right) ^{2}.  \label{ML-Hamiltonian}
\end{equation}%
In Figure 1(a) we show the response of the PDM system of (\ref{x(t)-MLPDM}) to the values of the damping ratio $\eta$. It is clear that our PDM system of (\ref{ML-m(x) equation of motion}) (a Mathews-Lakshmanan type PDM) inherits the traditional/standard constant mass behaviour as it evolves with time. We observe that in the \emph{under-damped} case, $\eta<1$ (blue curve), the system oscillates with a slightly different frequency than the \emph{undamped} case, $\eta=0$ (red curve), with its amplitude decreasing to zero. Moreover, for the \emph{critical-damping} case, $\eta=1$ (green curve), the system exponentially returns to the steady state (i.e., $x(t)=0$) faster than the \emph{over-damping} case, $\eta>1$ (purple curve). In Figures 1(b) and 1(c), we show the effects of the amplitude $A$ and the mass parameter $\lambda$, respectively, on the behaviour of the \emph{under-damping} case.  Both figures clearly indicate that the frequencies of oscillations are not affected by the values of the amplitude or the mass parameter. Such a behaviour  is a signature for isochronuous damped harmonic oscillators. In Figures 1(d), 1(e), and 1(f), it is obvious that the PDM phase-space trajectories of the \emph{under-damped} oscillators shrinks as time goes on and tends to wards the origin (the steady/equilibrium state)
\subsubsection{A coordinate deformation with singularity: $Q\left( x\right)=1/\left( 1-\lambda x\right) $}

Such coordinate deformation would result in a PDM-function%
\begin{equation}
m\left( x\right) =\frac{1}{4}\frac{\left( 2-\lambda x\right) ^{2}}{\left(
1-\lambda x\right) ^{3}}.  \label{singular PDM}
\end{equation}%
In this case, equations (\ref{point-transformation1}) and (\ref{q-adopted}) would result%
\begin{equation}
q^{2}=\frac{x^{2}}{1-\lambda x}\Longrightarrow x=x_{\pm}(t)=\frac{q}{2}\left( -\lambda
q\pm \sqrt{\lambda ^{2}q^{2}+4}\right) ;\,q=Ae^{-\omega _{\circ }\eta
t}\cosh \left( \beta t\right) ,  \label{solution for singular PDM}
\end{equation}%
as the exact solution for the one-dimensional form of (\ref{EL-transformed}) that reads the dynamical equation of motion%
\begin{equation}
\ddot{x}-\frac{\lambda \left( \lambda x-4\right) }{2\left( \lambda
x-1\right) \left( \lambda x-2\right) }\dot{x}^{2}+2\eta \omega _{\circ }\dot{%
x}+\frac{2\left( \lambda x-1\right) }{\left( \lambda x-2\right) }\omega
_{\circ }^{2}x=0.  \label{singular PDM eq of motion}
\end{equation}%
Which describes a PDM particle with a deformation in its velocity (\ref{singular PDM}) moving in the vicinity of a conservative PDM-deformed potential force field%
\begin{equation}
V\left( x\right) =\frac{\omega _{\circ }^{2}}{2\left( 1-\lambda x\right) }%
x^{2},  \label{singular PDM V}
\end{equation}%
and feels a non-conservative PDM-deformed Rayleigh dissipative force field%
\begin{equation}
\mathcal{R}\left( x,\dot{x}\right) =\frac{1}{4}\frac{\left( \lambda x-2\right)
^{2}}{\left( 1-\lambda x\right) ^{3}}b\dot{x}^{2}.  \label{singular PDM G}
\end{equation}%
Then the standard PDM-Lagrangian and PDM-Hamiltonian that describe this PDM-particle in (\ref{singular PDM eq of motion}) read%
\begin{equation}
L=\frac{1}{8}\frac{\left( 2-\lambda x\right) ^{2}}{\left( 1-\lambda x\right)
^{3}}\dot{x}^{2}-\frac{\omega _{\circ }^{2}}{2\left( 1-\lambda x\right) }%
x^{2}\Rightarrow H=\frac{2 \ \left(1-\lambda x\right)^{3}}{\left( 2-\lambda x\right) ^{2}}\ p_{_{x}}^2+\frac{\omega _{\circ }^{2}}{2\left(
1-\lambda x\right) }x^{2}  \label{singular PDM-L-H}
\end{equation}%
In Figures 2(a) and 2(b) we observe similar trends of response of the PDM system of (\ref{solution for singular PDM}) to the values of the damping ratio $\eta$.  The PDM phase-space trajectories of the PDM \emph{under-damped} oscillators, Figures 2(c) and 2(d), are observed to shrink with time.
\begin{figure}[h!]  
\centering
\includegraphics[width=0.4\textwidth]{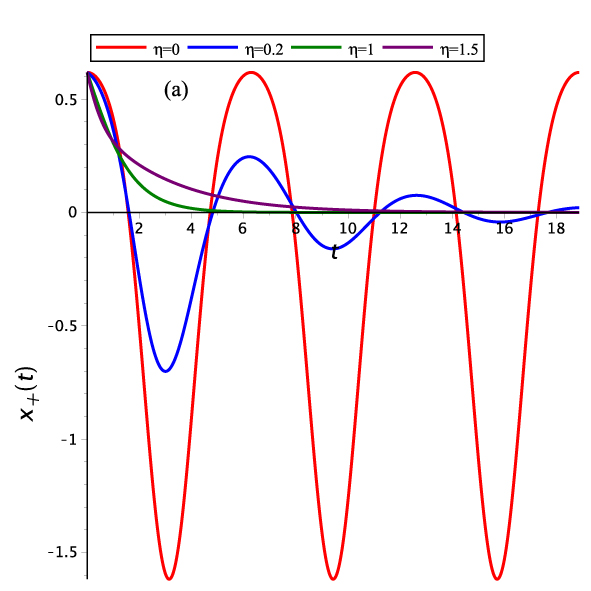} 
\includegraphics[width=0.4\textwidth]{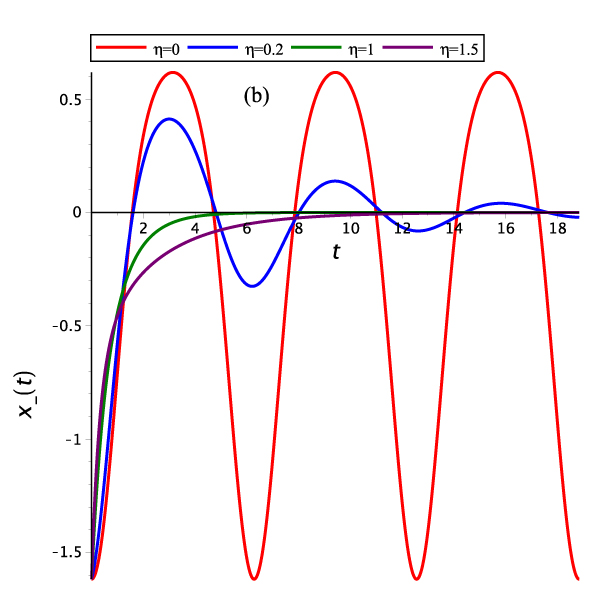}
\includegraphics[width=0.4\textwidth]{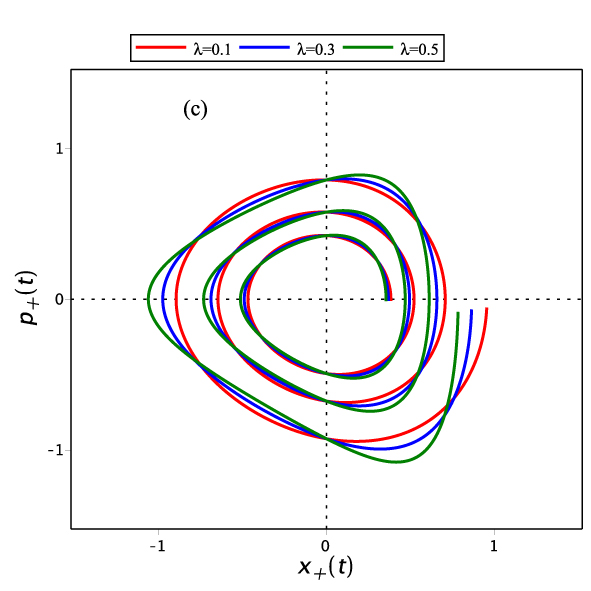}
\includegraphics[width=0.4\textwidth]{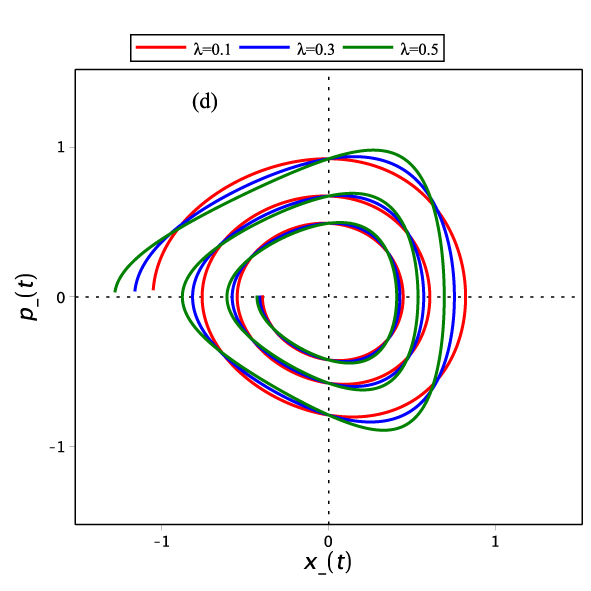}
\caption{\small 
{ For the PDM (\ref{singular PDM}) we plot the coordinates $x_{\pm}(t)$ of the solution (\ref{solution for singular PDM})  at $A=1, \omega=1, \lambda=1$, and $\eta=0.0, 0.2, 1.0, 1.5$, where (a) is for $x_{+}(t)$ and (b) for $x_{-}(t)$. The phase trajectories at $A=1, \omega=1, \eta=0.05$ (under-damped) are plotted for (c) $p_{+}(t)$ vs $x_{+}(t)$  and for (d) $p_{-}(t)$ vs $x_{-}(t)$ at $\lambda=0.1, 0.3, 0.5$.}}
\label{fig2}
\end{figure}
\begin{figure}[h!]  
\centering
\includegraphics[width=0.4\textwidth]{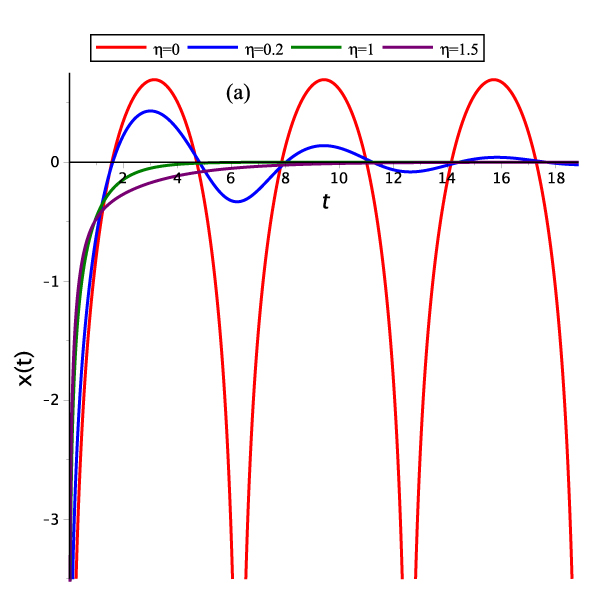} 
\includegraphics[width=0.4\textwidth]{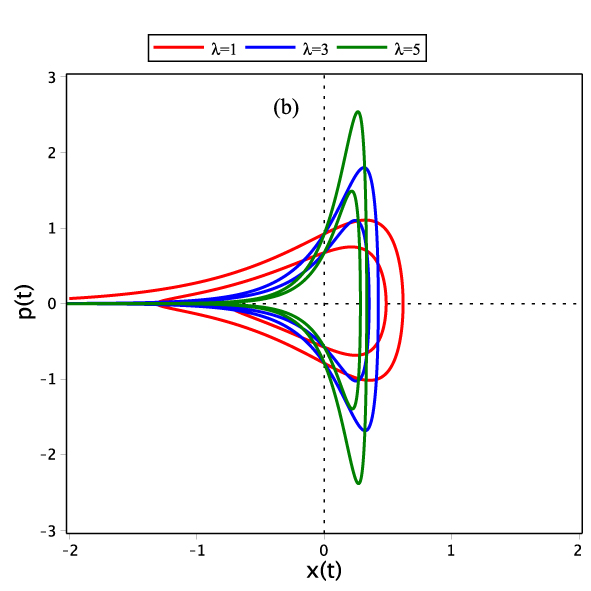}
\caption{\small 
{ For the damped Morse-type PDM-oscillator with $m(x)=e^{2\lambda x}$ we plot (a) the coordinate $x(t)$ of (\ref{damped PDM Morse solution}) as it evolves with time at $A=1, \omega=1, \lambda=1$ and $\eta=0.0, 0.2, 1.0, 1.5$. The corresponding phase trajectories $p(t)$ vs $x(t)$ at $A=1, \omega=1, \eta=0.05$ (under-damped) are plotted in (d) for $\lambda=1, 3, 5$. 
}}
\label{fig3}
\end{figure}

\subsubsection{A damped Morse-type PDM-oscillator: $m\left( x\right) =e^{2\lambda x}$}

A velocity deformation of an exponential form $m\left( x\right) =e^{2\lambda
x}$ would imply that%
\begin{equation}
Q\left( x\right) =\frac{e^{2\lambda x}}{\lambda ^{2}x^{2}}\left(
1-e^{-\lambda x}\right) ^{2}\Rightarrow q=\frac{1}{\lambda }\left(
e^{\lambda x}-1\right) =Ae^{-\omega _{\circ }\eta t}\cosh \left( \beta
t\right) .  \label{Morse q}
\end{equation}%
Which results%
\begin{equation}
x=\frac{1}{\lambda }\ln \left( \lambda q+1\right) \Longrightarrow x\left(
t\right) =\frac{1}{\lambda }\ln \left( \lambda Ae^{-\omega _{\circ }\eta
t}\cosh \left( \beta t\right) \right) ,  \label{damped PDM Morse solution}
\end{equation}%
as an exact solution that satisfies the corresponding equation of motion%
\begin{equation}
\ddot{x}+\lambda \dot{x}^{2}+2\eta \omega _{\circ }\dot{x}+\frac{\omega
_{\circ }^{2}}{\lambda }\left( 1-e^{-\lambda x}\right) =0.
\label{Damped PDM Morse}
\end{equation}%
This equation belongs to a PDM-particle moving under the influence of a conservative PDM-deformed potential force field%
\begin{equation}
V\left( x\right) =\frac{1}{2}\omega _{\circ }^{2}\frac{e^{2\lambda x}}{%
\lambda ^{2}}\left( 1-e^{-\lambda x}\right) ^{2},  \label{PDM Morse V}
\end{equation}%
and a non-conservative PDM-deformed Rayleigh dissipative force field%
\begin{equation}
\mathcal{R}\left(x, \dot{x}\right) =be^{2\lambda x}\dot{x}^{2}.
\label{Morse PDM G}
\end{equation}%
Moreover, the standard PDM-Lagrangian and PDM-Hamiltonian that describe this PDM-particle are given by%
\begin{equation}
L=\frac{1}{2}e^{2\lambda x}\dot{x}^{2}-\frac{1}{2}\omega _{\circ }^{2}\frac{%
e^{2\lambda x}}{\lambda ^{2}}\left( 1-e^{-\lambda x}\right)
^{2}\Longrightarrow H=\frac{1}{2}e^{-2\lambda x} \ p_{_{x}}^{2}+\frac{1}{2}\omega
_{\circ }^{2}\frac{e^{2\lambda x}}{\lambda ^{2}}\left( 1-e^{-\lambda
x}\right) ^{2}.  \label{Morse PDM-L-H}
\end{equation}%
In Fig. 3(a), we observe similar trends of response of the PDM system of (\ref{damped PDM Morse solution}) to the values of the  damping ratio,  as those of the examples reported above. So are the phase-space trajectories, Fig. 3(b).
\begin{figure}[h!]  
\centering
\includegraphics[width=0.4\textwidth]{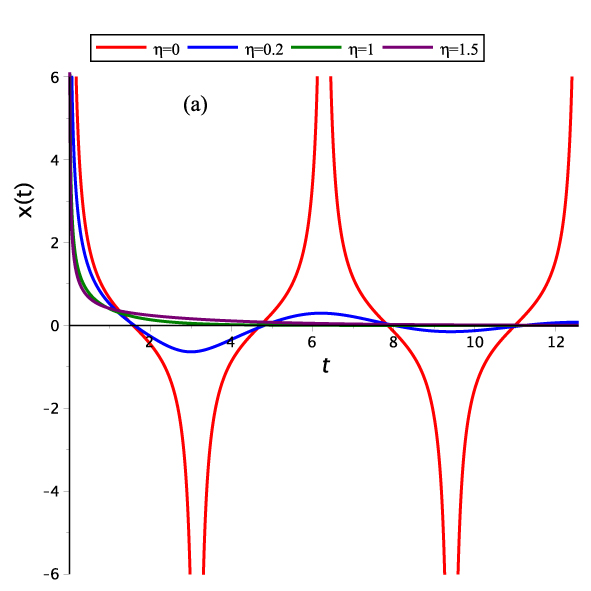} 
\includegraphics[width=0.4\textwidth]{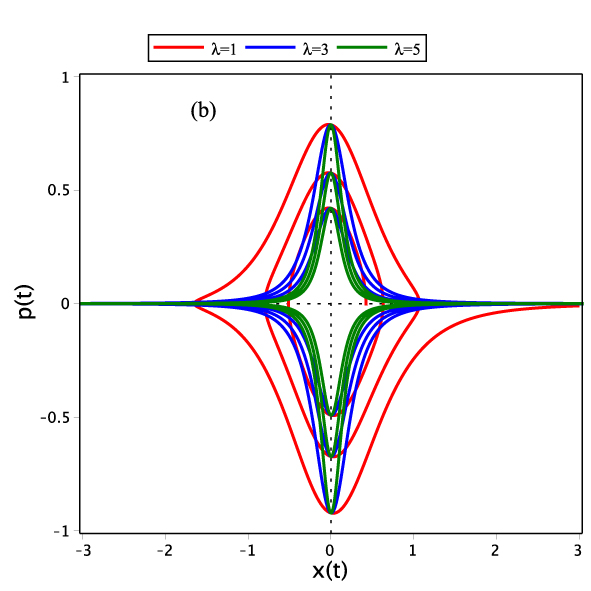}
\caption{\small 
{For the PDM (\ref{m-Q ML}) we plot in (a) the $ith$ coordinate $x_{i}(t)$ in (\ref{nD-deformed coordinate}) as it evolves with time at $A=1, \omega=1, \lambda=1$ for $\eta=0.0, 0.2, 1.0, 1.5$. The corresponding phase trajectories $p_{i}(t)$ vs $x_{i}(t)$ with $A=1, \omega=1, \eta=0.05$ (under-damped) are shown in (b) for $\lambda=1, 3, 5$.
}}
\label{fig4}
\end{figure}
\subsection{$n$-dimensional PDM damped oscillators}

\subsubsection{Coordinates deformed by $Q\left( r\right) =1/\left( 1+\lambda ^{2}r^{2}\right) $}

A deformation in the coordinates in the form of%
\begin{equation}
Q\left( r\right) =\frac{1}{1+\lambda ^{2}r^{2}};\,r^{2}=\sum%
\limits_{j=1}^{n}x_{j}^{2},  \label{Q-m ML}
\end{equation}%
would result a PDM of the form%
\begin{equation}
m\left( r\right) =\frac{1}{\left( 1+\lambda ^{2}r^{2}\right) ^{3}}.
\label{m-Q ML}
\end{equation}%
Using the point transformation (\ref{point-transformation1}) would lead, with $q^{2}=\sum\limits_{j=1}^{n}q_{j}^{2},$ to%
\begin{equation*}
q^{2}=Q\left( r\right) r^{2}\Rightarrow r^{2}=\frac{q^{2}}{1-\lambda
^{2}q^{2}}\Rightarrow \mathbf{r}=\frac{1}{1-\lambda ^{2}q^{2}}\mathbf{%
q\Rightarrow }x_{i}=\frac{1}{1-\lambda ^{2}q^{2}}q_{i}.
\end{equation*}%
Consequently, with $q_{i}=A_{i}\,e^{-\omega _{\circ }\eta t}\cosh \left(
\beta t\right) $ and $A^{2}=\sum\limits_{j=1}^{n}A_{j}^{2}$,%
\begin{equation}
x_{i}\left( t\right) =\frac{A_{i}\,\cosh \left( \beta t\right) }{\sqrt{%
e^{2\omega _{\circ }\eta t}-\lambda ^{2}A^{2}\cosh ^{2}\left( \beta t\right)
}} ; \lambda<\frac{1}{A},  \label{nD-deformed coordinate}
\end{equation}%
which serve as the $n$-dimensional exact solutions for the $n$ dynamical standard PDM damped oscillators equations%
\begin{equation}
\ddot{x}_{i}-\frac{3\lambda ^{2}}{1+\lambda ^{2}r^{2}}\left(
\sum\limits_{j=1}^{n}\dot{x}_{_{j}}^{2}\right) x_{i}+2\eta \omega _{\circ }%
\dot{x}_{i}+\omega _{\circ }^{2}\left( 1+\lambda ^{2}r^{2}\right) \,x_{i}=0.
\label{nD damped 1}
\end{equation}%
These dynamical equations belong to a PDM particle, $m\left( r\right)
=\left( 1+\lambda ^{2}r^{2}\right) ^{-3}$, described by the $n$-dimensional
PDM -Lagrangian and PDM-Hamiltonian%
\begin{equation}
L=\frac{1}{2\left( 1+\lambda ^{2}r^{2}\right) ^{3}}\sum\limits_{j=1}^{n}\dot{%
x}_{_{j}}^{2}-\frac{\omega _{\circ }^{2}}{2\left( 1+\lambda ^{2}r^{2}\right) 
}\sum\limits_{j=1}^{n}x_{_{j}}^{2}\Longrightarrow H=\left(
1+\lambda ^{2}r^{2}\right) ^{3}\frac{p^2}{2}+\frac{\omega _{\circ }^{2}}{2\left( 1+\lambda ^{2}r^{2}\right) }%
\sum\limits_{j=1}^{n}x_{_{j}}^{2}  \label{L-H nD Q(x)}
\end{equation}%
and feels a non-conservative $n$-dimensional PDM-deformed Rayleigh dissipative force field%
\begin{equation}
\mathcal{R}\left( \mathbf{r},\mathbf{\dot{r}}\right) =\frac{b}{\left( 1+\lambda
^{2}r^{2}\right) ^{3}}\sum\limits_{j=1}^{n}\dot{x}_{_{j}}^{2}.
\label{nD-G Q(x)}
\end{equation}%
\begin{figure}[h!]  
\centering
\includegraphics[width=0.3\textwidth]{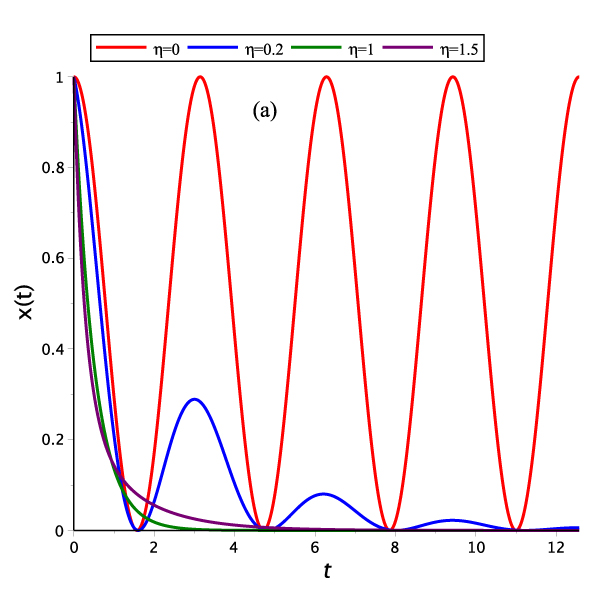} 
\includegraphics[width=0.3\textwidth]{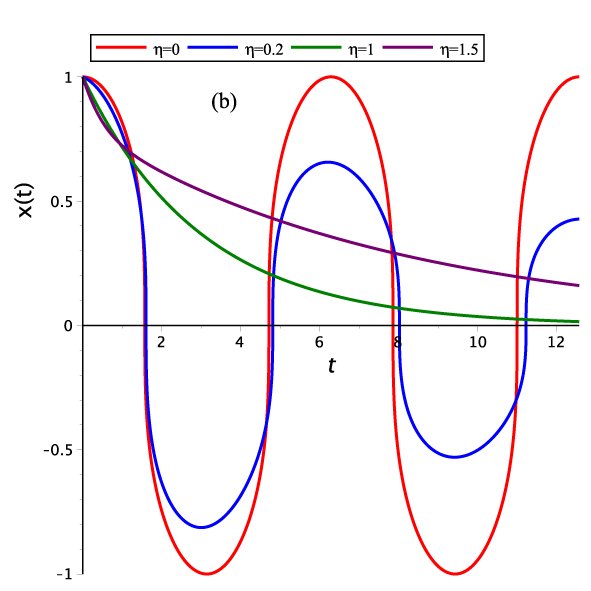}
\includegraphics[width=0.3\textwidth]{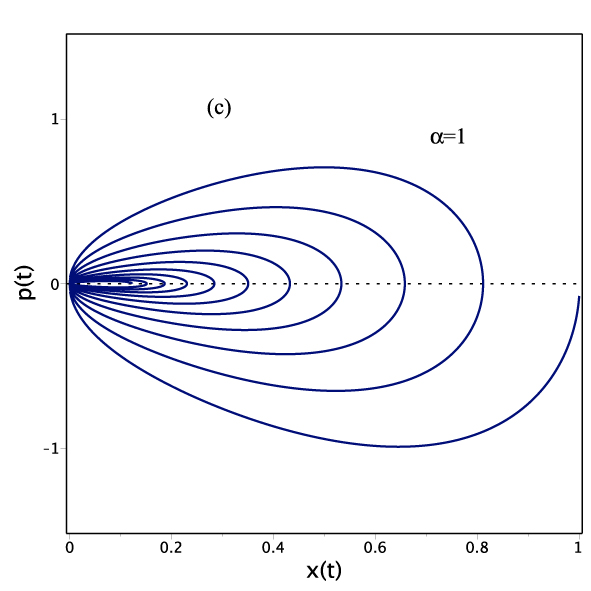}
\includegraphics[width=0.3\textwidth]{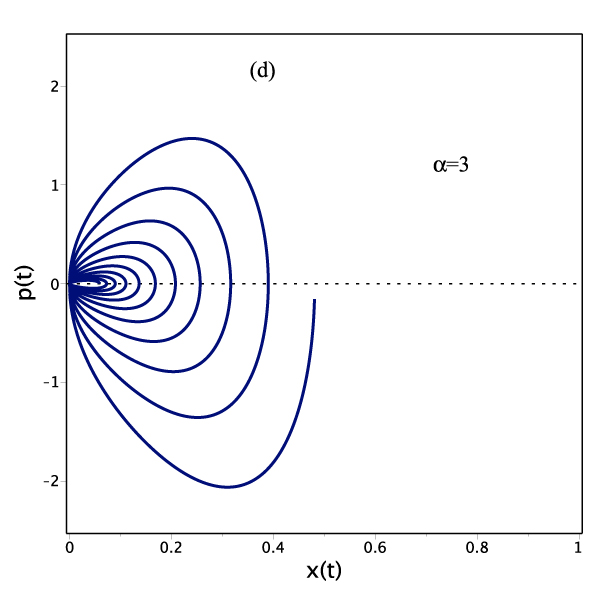}
\includegraphics[width=0.3\textwidth]{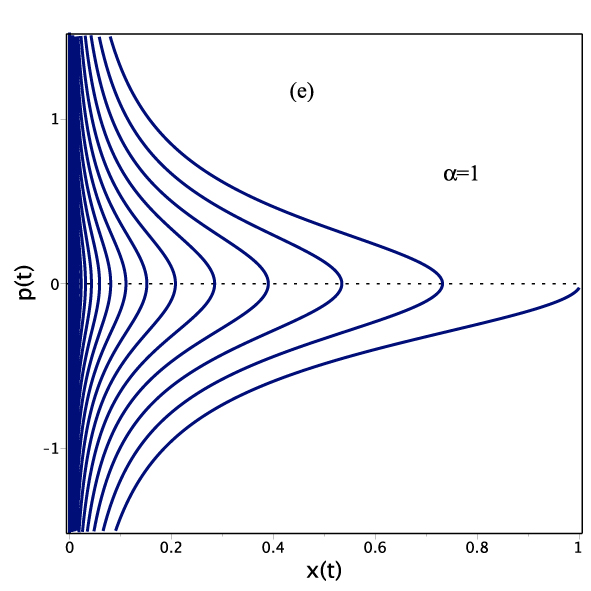}
\includegraphics[width=0.3\textwidth]{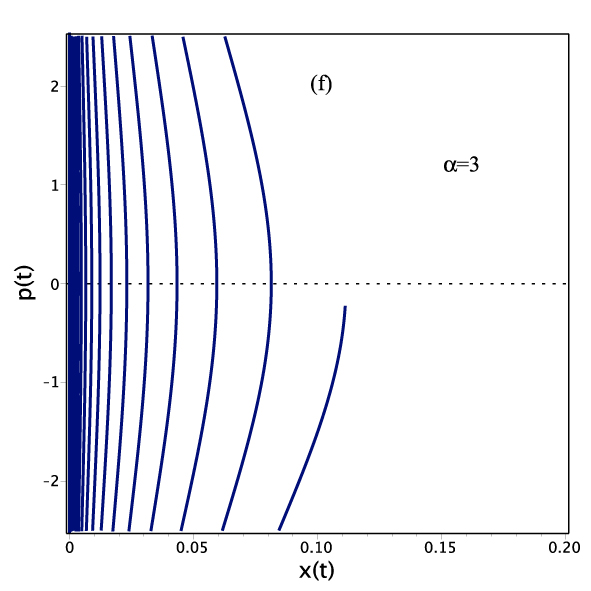}
\caption{\small 
{ For the power-law type PDM of (\ref{power law m}) we plot the coordinate $x(t)$ in (\ref{x(t) power law}) at $A=1, \omega=1, \alpha=1$ and $\eta=0.0, 0.2, 1.0, 1.5$ in (a) for $\sigma=-0.5$ and in (b) for $\sigma=2$. The effect of the PDM parameter $\alpha$ on the phase trajectories are plotted at $A=1, \omega=1, \eta=0.05$ (under-damped), where for $\sigma =0.5$ we plot in (c) for $\alpha=1$ and in (d) for $\alpha=3$. For $\sigma=-0.5$ we plot (e) for $\alpha=1$ and (f) for $\alpha=3$.
}}
\label{fig5}
\end{figure}
In Figure 4(a), for the PDM in (\ref{m-Q ML}) we show the system  response to the damping ratio $\eta$ and plot the corresponding phase-space trajectories at different PDM parameter $\lambda$-values. Similar trends are observed for both the $ith$ coordinate (\ref{nD-deformed coordinate}) and the corresponding phase space trajectories.%

\subsubsection{ Power-law type PDM deformed damped oscillators}

A power-law  type coordinates deformation in the form of%
\begin{equation}
Q\left( r\right) =\alpha \,r^{\sigma }  \label{power law Q}
\end{equation}%
would imply a PDM function%
\begin{equation}
m\left( r\right) =\alpha ^{2}\left( \sigma +1\right) ^{2}r^{2\sigma }
\label{power law m}
\end{equation}%
In this case, with $q_{i}=A_{i}\,e^{-\omega _{\circ }\eta t}\cosh \left(\beta t\right) $, $q^{2}=\sum\limits_{j=1}^{n}q_{j}^{2}$ and $A^{2}=\sum\limits_{j=1}^{n}A_{j}^{2}$,%
\begin{equation}
\mathbf{q}=\alpha \,r^{\sigma }\mathbf{r}\Longleftrightarrow \mathbf{r}%
=\left( \frac{q^{-\sigma }}{\alpha }\right) ^{1/\left( \sigma +1\right) }%
\mathbf{q}\Longleftrightarrow x_{i}\left( t\right) =\left( \frac{q^{-\sigma }%
}{\alpha }\right) ^{1/\left( \sigma +1\right) }A_{i}\,e^{-\omega _{\circ
}\eta t}\cosh \left( \beta t\right) ,  \label{x(t) power law}
\end{equation}%
which satisfies the $n$-equations of motion for the damped PDM oscillators%
\begin{equation}
\ddot{x}_{i}+\frac{\sigma }{r^{2}}\left( \sum\limits_{j=1}^{n}\dot{x}%
_{_{j}}^{2}\right) x_{i}+2\eta \omega _{\circ }\dot{x}_{i}+\frac{\omega
_{\circ }^{2}}{\sigma +1}\,x_{i}=0.  \label{power law eq of motion}
\end{equation}%
These $n$-PDM dynamical equations of motion for a PDM particle $m\left(
r\right) =\alpha ^{2}\left( \sigma +1\right) ^{2}r^{2\sigma }$ described by the $n$-dimensional standard PDM-Lagrangian and PDM-Hamiltonian%
\begin{equation}
L=\frac{\alpha ^{2}\left( \sigma +1\right) ^{2}r^{2\sigma }}{2}\left(
\sum\limits_{j=1}^{n}\dot{x}_{_{j}}^{2}\right) -\frac{\omega _{\circ }^{2}}{2%
}\alpha \,r^{\sigma }\sum\limits_{j=1}^{n}x_{_{j}}^{2}\Longleftrightarrow H=%
\frac{\alpha ^{-2}\left( \sigma +1\right) ^{-2}r^{-2\sigma }}{2} \ p^2+\frac{\omega _{\circ }^{2}}{2%
}\alpha \,r^{\sigma }\sum\limits_{j=1}^{n}x_{_{j}}^{2}  \label{power law L-H}
\end{equation}%
and feels a non-conservative $n$-dimensional PDM-deformed Rayleigh dissipative force field%
\begin{equation}
\mathcal{R}\left(\mathbf{r}, \mathbf{\dot{r}}\right) =b\alpha ^{2}\left( \sigma
+1\right) ^{2}r^{2\sigma }\sum\limits_{j=1}^{n}\dot{x}_{_{j}}^{2}.
\label{power law G}
\end{equation}%
In Figures 5(a) and 5(b) we show the dynamical system (\ref{x(t) power law})  response, for the PDM in (\ref{power law m}), to the damping ratio for $\sigma=-0.5$ and $\sigma=2$ , respectively. The phase-space trajectories at $\sigma=0.5$ are shown in 5(c) for the PDM parametric value $\alpha=1$, and in 5(d) for $\alpha=3$. At $\sigma=-0.5$ we use in 5(e) the PDM parametric value $\alpha=1$, and in 5(f)  $\alpha=3$.
\section{Concluding Remarks}

In this work, we have considered standard PDM-Lagrangians/PDM-Hamiltonians, for which the \emph{gain-loss balance} correlation between the kinetic and potential energies is preserved to secure conservation of the total energy (i.e., $L=T-V$, $H=T+V$, and $dH/dt=dE/dt=0$). Within such standard settings, we have discussed and reported on the $n$-dimensional PDM damped harmonic oscillators subjected to a conservative PDM-harmonic oscillator force field (\ref{PDM-oscillator}) and a non-conservative PDM-Rayleigh dissipative force field (\ref{PDM-Rayleigh}). We have used the $n$-dimensional point canonical transformation (\ref{point-transformation1}) and (\ref{point-transformation2}), along with the total vector \emph{Newtonian invariance amendment} in (\ref{Newtonian invariance}) (which has been recently introduced by Mustafa \cite {Mustafa Phys.Scr. 2020,Mustafa arXiv}),  to facilitate the linearizability of their $n$-PDM dynamical equations (\ref{PDM damped oscillator}) into some $n$-linear DHOs' dynamical equations (\ref{q-equation of motion}) for constant mass setting. Consequently, the well know exact solutions for the linear DHOs  (\ref{q-adopted}) are mapped, with ease, onto the exact solutions for PDM DHOs. This is documented in the set of one-dimensional and the set of $n$-dimensional PDM-DHO illustrative examples of section III. Their phase-space trajectories are also reported in Figures 1-5. To the best of our knowledge, neither the current methodical approach nor the current results have been reported elsewhere.

In the light of our experience above, our observations are in order.

In connection with the prominent nonlinear dynamical systems of Li\'{e}nard-type (c.f., e.g., Ref\cite{Lak-Chand 2013}), we have effectively shown that a mixed Li\'{e}nard-type differential equation%
\begin{equation*}
\ddot{x}+f\left( x\right) \dot{x}^{2}+h\left( x\right) \dot{x}+g\left( x\right) =0;  f\left( x\right)=\frac{m^{\prime }\left( x\right)}{2m\left( x\right)}, h\left( x\right) =2\eta\omega _{\circ }, g\left( x\right) =\sqrt{\frac{Q\left( x\right)}{m\left(x\right)} }\omega ^{2}x
\end{equation*}%
is linearizable into a linear Li\'{e}nard-type differential equation%
\begin{equation*}
\ddot{q}+\tilde{h}\left( q\right) \dot{q}+\tilde{g}\left( q\right) =0;  \tilde{h}\left( q\right) =2\eta \omega , \tilde{g}\left( q\right) =\omega ^{2}q.
\end{equation*}%
Moreover, if the damping ratio is switched off, $\eta=0$, then $h(x)=0$ and the mixed Li\'{e}nard-type differential equation reduces into a quadratic  Li\'{e}nard-type. Consequently, the later is linearizable into a simple harmonic oscillator one. All of which admit exact solvability within the lines discussed above. The current methodical proposal offers not only a user friendly and a straightforward approach that allows transition between the three   Li\'{e}nard-type dynamical equations, but also provides their PDM damped and/or undamped harmonic oscillator counterparts and facilitates their exact solvability.

Yet, it should be made clear that, the Lagrangian/Hamiltonian structures for dissipative damped harmonic oscillator systems used early on by Pradeep et al. \cite{Pradeep 2009}, Chandrasekar et al. \cite{Chand-Lak 2007} and the related comment of Bender et al. \cite{Carl Bender 2016} do not belong to the class of the standard textbook PDM-Lagrangians/PDM-Hamiltonians used in the current methodical proposal. The two completely different approaches should not be confused with each other, therefore.

On the feasibly viable applicability of the above methodical proposal, it would be interesting to study particles subjected to drag (of fluid, say) and random forces (effect of Brownian motion) to yield Langevin equation. Where the classical description is connected with a probabilistic one, which would suggest that the Fokker-Plank equation (PFE) inhomogeneous medium with variable diffusion PDM-coefficient is equivalent with deformed FPE with constant mass and constant diffusion coefficient. For more details on this issue the reader may refere to  (c.f., e.g. da Costa et al. \cite{da Costa1 2020}). Among other possible applications are  the Killing vector fields and Noether momenta approach for quantum PDM Hamiltonians of Cari$\tilde{n}$ena et al. \cite{Carinena Ranada Sant 2017},  the generation of integrable nonlinear dynamical systems of Lakshmanan-Chandrasekar \cite{Lak-Chand 2013}, etc.


\end{document}